\documentclass[aps,prd,secnumarabic,nobibnotes,twocolumn,superscriptaddress]{revtex4-1}
\usepackage{amsfonts}
\usepackage{mathrsfs}
\usepackage{amsmath}% needed for subequations
\usepackage{color}
\usepackage{natbib}
\usepackage{graphicx}
\usepackage{bm}% bold maths
\usepackage{amssymb}
\usepackage{xspace}
\usepackage{epstopdf}
\usepackage{dcolumn}
\usepackage{multirow}
\usepackage[colorlinks=true, letterpaper=true, pdfstartview=FitV, linkcolor=blue, citecolor=blue, urlcolor=blue]{hyperref}
\usepackage{wrapfig}

\makeatletter

\newcommand{\Rmnum}[1]{\expandafter\@slowromancap\romannumeral #1@}
\makeatother

\begin{document}
\title{Fully spin-polarized nodal chain state in half metal LiV$_2$O$_4$}

\author{Haopeng Zhang}
\affiliation{School of Materials Science and Engineering, Hebei University of Technology, Tianjin 300130, China.}

\author{Xiaoming Zhang}
\email{zhangxiaoming87@hebut.edu.cn}
\affiliation{School of Materials Science and Engineering, Hebei University of Technology, Tianjin 300130, China.}

\author{Ying Liu}
\affiliation{School of Materials Science and Engineering, Hebei University of Technology, Tianjin 300130, China.}
\affiliation{Research Laboratory for Quantum Materials, Singapore University of Technology and Design, Singapore 487372, Singapore}

\author{Xuefang Dai}
\affiliation{School of Materials Science and Engineering, Hebei University of Technology, Tianjin 300130, China.}

\author{Guang Chen}
\affiliation{BaodingFengfanRisingBatterySeparatorCo.,Ltd., Baoding 071052, Hebei province, People's Republic of China.}

\author{Guodong Liu}
\email{gdliu1978@126.com}
\affiliation{School of Materials Science and Engineering, Hebei University of Technology, Tianjin 300130, China.}

\begin{abstract}
Nodal-chain fermions, as novel topological states of matter, have been hotly discussed in nonmagnetic materials. Here, by using first-principles calculations and symmetry analysis, we propose the realization of fully spin-polarized nodal chain in the half-metal state of LiV$_2$O$_4$ compound. The material naturally shows a ferromagnetic ground state, and takes on a half-metal band structure with only the bands from the spin-up channel present near the Fermi level. The spin-up bands cross with each other, which form two types of nodal loops. These nodal loops arise from band inversion and are under the protection of the glide mirror symmetries. Remarkably, we find the nodal loops conjunct with each other and form chain-like nodal structure. Correspondingly, the $\omega$-shaped surface states are also fully spin-polarized. The fully spin-polarized nodal chain identified here has not been proposed in realistic materials before. An effective model is constructed to describe the nature of nodal chain. The effects of the electron correlation, the lattice strains, and the spin-orbit coupling are discussed. The fully spin-polarized bulk nodal-chain and the associated nontrivial surface states for a half-metal may open novel applications in spintronics.
\end{abstract}
\maketitle

%%%%%%% Main text %%%%%%%%%%%%%%%%%%%%%
\section{INTRODUCTION}
Topological semimetals have attracted tremendous attention recently, because they not only show potential applications in various fields but also provide a feasible bride to investigate the novel characteristics of high-energy particles~\cite{1,2,3,4,5,6,7,8,9}. In topological semimetals, the crossings between the valence and conduction bands can show different dimensionalities; thereby form different types of topological semimetals with hosting nodal points (such as Dirac/Weyl points)~\cite{10,11,12,13,14,15}, nodal lines~\cite{16,17,18,19}, and nodal surfaces~\cite{20,21,22,23}. Nodal lines have many nodal structures. For example, one single nodal line can either traverse the whole Brillouin zone or form a closed loop without penetrating the BZ, as distinguished by the $Z^3$ index~\cite{24}. For another example, nodal line can be classified as type-I, type-II, critical-type and hybrid line according to the slope of crossing bands~\cite{24,25,26,27}. In addition, when multiple nodal lines coexist in the Brillouin zone, they can form different configurations such as crossing loops, nodal net, nodal box, Hopf link, and nodal chain~\cite{28,29,30,31,32,33,34,35,36,37}. Recently, nodal-chain semimetal has been hotly discussed~\cite{33,34,38,39,40,41,42,43,44}. They contain a chain of connected loops in the momentum space and the connecting points are under the protection of special symmetries.

Comparing with the nonmagnetic counterparts, magnetic topological semimetals have currently attracted increasing attention because they are special in several aspects. From the fundamental physic point of view, the time-reversal symmetry (TRS) is broken in magnetic system. For example, TRS-breaking Weyl semimetals are distinct and in some degree simper than the inversion-symmetry-breaking ones such as the nonzero anomalous Hall conductivity, as well as the potential of containing the minimum one pair of Weyl points in the Brillouin zone ~\cite{45,46,47,48,49,50,51,52}. In addition, in magnetic system, the magnetic symmetry highly relies on the magnetization direction. As the result, novel topological phase transition may happen by shifting the magnetic symmetry, controlled by external magnetic field~\cite{53,54,55,56}. Magnetic topological semimetals are also desirable from the application point of view, because they are promising to be applied in spintronic devices. A series of magnetic topological semimetals with variable fermionic states have been proposed~\cite{57,58,59,60}. In particular, it will be the most desirable if the topological fermions have a 100$\%$ spin-polarization, namely topological half metals. Compounds HgCr$_2$Se$_4$~\cite{61}, and Co$_3$Se$_2$S$_2$~\cite{45} are examples for Dirac/Weyl half metals; Li$_3$(FeO$_3$)$_2$ compound~\cite{62}, tetragonal $\beta$-V$_2$PO$_5$~\cite{53}, and MnN monolayer~\cite{63} are examples for nodal loop half metals;  CsCrX$_3$ (X = Cl, Br, I) compounds are examples for nodal surface half metals~\cite{23}. Unfortunately, we have not retrieved reports on nodal chain half metals. Previously, nodal chains are mostly proposed in nonmagnetic materials ~\cite{33,38,39,40,41,42,43,44}. Heusler compound Co$_2$MnGa is almost the only example for magnetic nodal chain semimetal with hosting multiple types of nodal chains~\cite{34}. Unfortunately, this material is not a half metal, and the transports for nodal chains are not fully spin-polarized. Thus, it is highly desirable to explore nodal chain half metal with hosting fully spin-polarized nodal-chain fermion.

In current work, we report a spinel compound LiV$_2$O$_4$ is one such nodal chain half metal. It naturally shows a FM ordering with an integer magnetic moment of 6.0 $\mu_B$. The material is a half metal, which shows an insulating band structure in the spin-down channel but a metallic one in the spin-up channel. Especially, the conduct and valance bands in the spin-up channel cross with each other, which generates the fully spin-polarized nodal chain. The nodal chain is made up from two types of nodal loops, which are protected by the glide mirror symmetries. The nodal chain is characterized with the $\omega$-shaped surface states. We built an effective model, and the model can well describe the mechanism of nodal chain. We further find the nodal chain half metal is very robust against the electron correlation effects and the lattice strain. The spin-orbit coupling (SOC) effect on the nodal chain is also discussed. The work suggests LiV$_2$O$_4$ compound is a good material platform to investigate the fundamental physics of nodal-chain fermion in ferromagnets.

\section{METHODS}

In this work, we perform the first-principles calculations by using the Vienna ab initio Simulation Package (VASP)~\cite{64}, based on density functional theory (DFT)~\cite{65}. The valence electron configurations of Li (2$s^1$), V (3$d^3$4$s^2$) and O(2$s^2$2$p^4$) are applied and the projector augmented wave method is adopted for the interaction between the valence electrons and the ionic core potentials~\cite{66}. During calculations, the cutoff energy is set as 600 eV. The Brillouin zone is sampled by a Monkhorst-Pack \emph{k}-mesh with size of 15$\times$15$\times$15. To optimize the lattice of LiV$_2$O$_4$, the force and energy convergence criteria are applied as 0.01 eV/\AA  and $10^{-6}$ eV, respectively. To account for the correlation effects for V element, the GGA + U method is applied to describe the Coulomb interaction~\cite{67}. The effective $U$ for V is set as 4 eV to investigate the topological band structure, and the conclusions will not change with $U$ values shifting from 0 eV to 10 eV. The irreducible representations of the electronic states are obtained by using the irvsp code~\cite{68}. The surface states are calculated by using the WANNIERTOOLS package~\cite{69}.

\section{ CRYSTAL AND MAGNETIC STRUCTURES}

The LiV$_2$O$_4$ compound is an existing material and has been synthesized as early as 1960~\cite{70}. It crystallizes in a normal spinel structure with the non-symmorphic space group \emph{Fd$\overline{3}$m }(No. 227). Figure 1(a) shows the crystal structure of LiV$_2$O$_4$ compound. In the crystal structure, the bonding between V and O atoms forms V-O$_6$ octahedral local structure; and the bonding between Li and O atoms forms Li-O$_4$ tetrahedral local structure. The cubic unit cell totally contains 56 atoms with 32 O atoms occupying the \emph{32e} (u, u, u), 16 V atoms at the \emph{16d} (0.5, 0.5, 0.5) and 8 Li atoms at the \emph{8a} (0.125, 0.125, 0.125) Wyckoff positions, respectively. One unit cell contains four primitive cells, and the primitive cell form of LiV$_2$O$_4$ is shown in Fig. 1(b). After lattice optimization, the ground lattice constants are found to be a=b=c= 8.465482 \AA, in good agreements with former computational and experimental ones~\cite{70,71,72}. In the following band structure calculations, we apply the optimized lattice structure. To be noted, the conclusions of this work will not change when the experimental structure is applied.

\begin{figure}
\includegraphics[width=8.8cm]{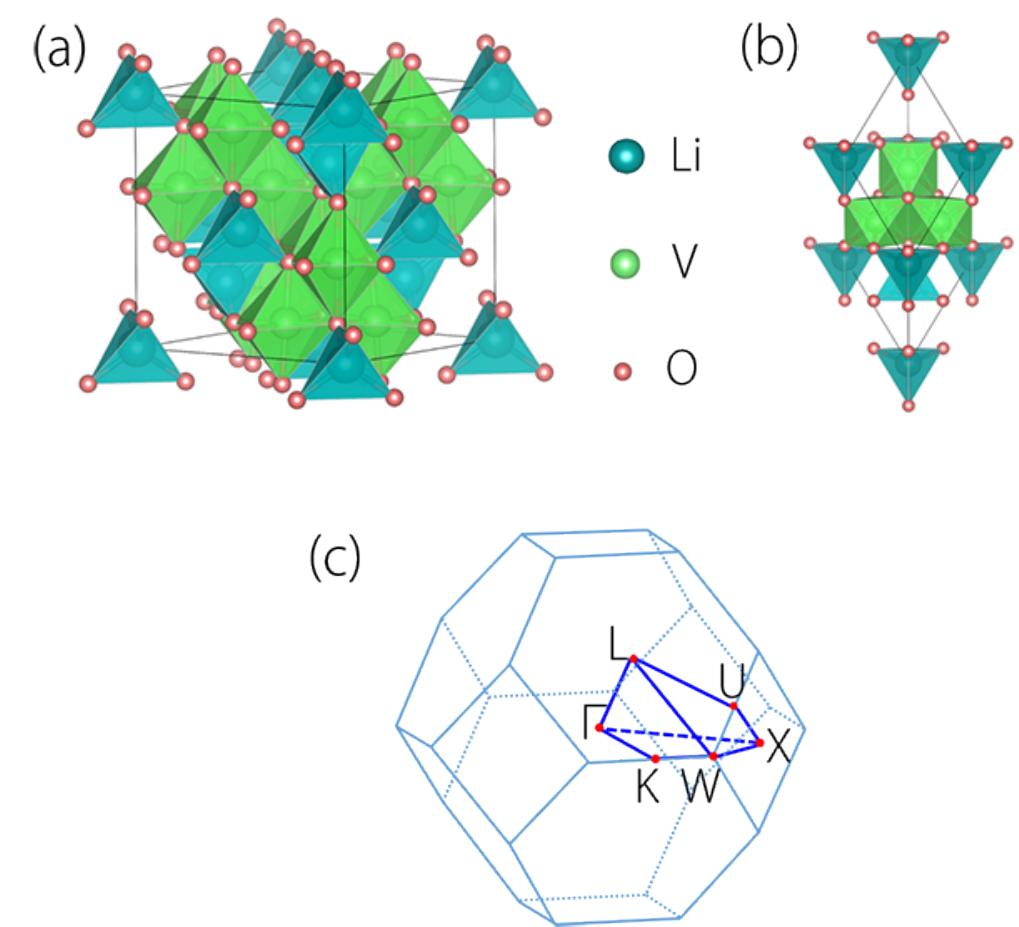}
\caption{(a) The conventional and (b) the primitive unit cell of crystal structure for LiV$_2$O$_4$. (c) The corresponding Brillouin zone with the considered high-symmetry paths.
\label{fig1}}
\end{figure}

Because of the unoccupied $d$ shells in transition-metal V element, it may carry magnetism moment in LiV$_2$O$_4$ compound. Here we determine the ground magnetic configuration by comparing the total energies among different magnetic states, which include ferromagnetic (FM), nonmagnetic (NM), and antiferromagnetic (AFM) states. In each magnetic states, we have considered three most potential magnetization directions in the cubic system including the [001], [110] and [111] directions. The obtained energies for all the magnetic configurations are summarized in Table I. It can be clearly found that, FM$_{[001]}$ has the lowest energy. Therefore, from our calculations, the ground magnetic configuration of LiV$_2$O$_4$ compound is FM, and the magnetic moment ordering is along in the [001] direction.

\begin{table}

% table caption is above the table

\caption{ Total energy E$_{tot}$ per unit cell (in eV, relative to that of the FM$_{001}$ ground state), as well as magnetic moment M (in units of $\mu_B$) per V atom. The values are calculated by the GGA+SOC method with U = 4.0 eV.}

\label{tab1}       % Give a unique label

% For LaTeX tables use

\begin{tabular}{lcccccccc}

\hline\noalign{\smallskip}

 & FM$_{001}$ & FM$_{110}$ & FM$_{111}$ & AFM$_{001}$ & AFM$_{110}$ & AFM$_{111}$ & NM \\

\noalign{\smallskip}\hline\noalign{\smallskip}

Energy/eV  & 0 & 0.243 & 0.187 & 0.199 & 0.296 & 0.285 & 5.441 \\

$M_x$/$\mu_B$ & 0.001 & 1.103 & 0.922 & 0.044 & 1.086 & 0.889 & 0 \\

$M_y$/$\mu_B$ & 0.001 & 1.122 & 0.912 & 0.034 & 1.089 & 0.888 & 0 \\

$M_z$/$\mu_B$  & 1.570 & 0.019 & 0.882 & 1.537 & 0.015 & 0.882 & 0 \\

\noalign{\smallskip}\hline

\end{tabular}

\end{table}

\section{Weyl nodal chain without SOC}

The electronic band structures of LiV$_2$O$_4$ compound in the absence of SOC are shown in Fig. 2 (a) and (b). The spin-resolved band structures show two features. First, the bands in the spin-up channel exhibit a metallic character with two bands crossing the Fermi level, whereas those in spin-down channel exhibit an insulating character with a big band gap of 3.27 eV. These results indicate LiV$_2$O$_4$ compound is a half metal, where the conduction electrons are fully spin-polarized. Second, in the spin-up channel, the two bands near the Fermi level cross with each other, and form several band crossings in the K-W, W-$\Gamma$, $\Gamma$-X and X-L paths. We will show later that these band crossings form nodal chains in the BZ. In Fig. 2 (a) and (b), we also show the total and projected density of states (TDOSs and PDOSs) of LiV$_2$O$_4$ compound. The TDOSs and PDOSs clearly show that, the electron states near the Fermi level are mostly contributed by the $d$ orbitals of V element.

\begin{figure}
\includegraphics[width=8.8cm]{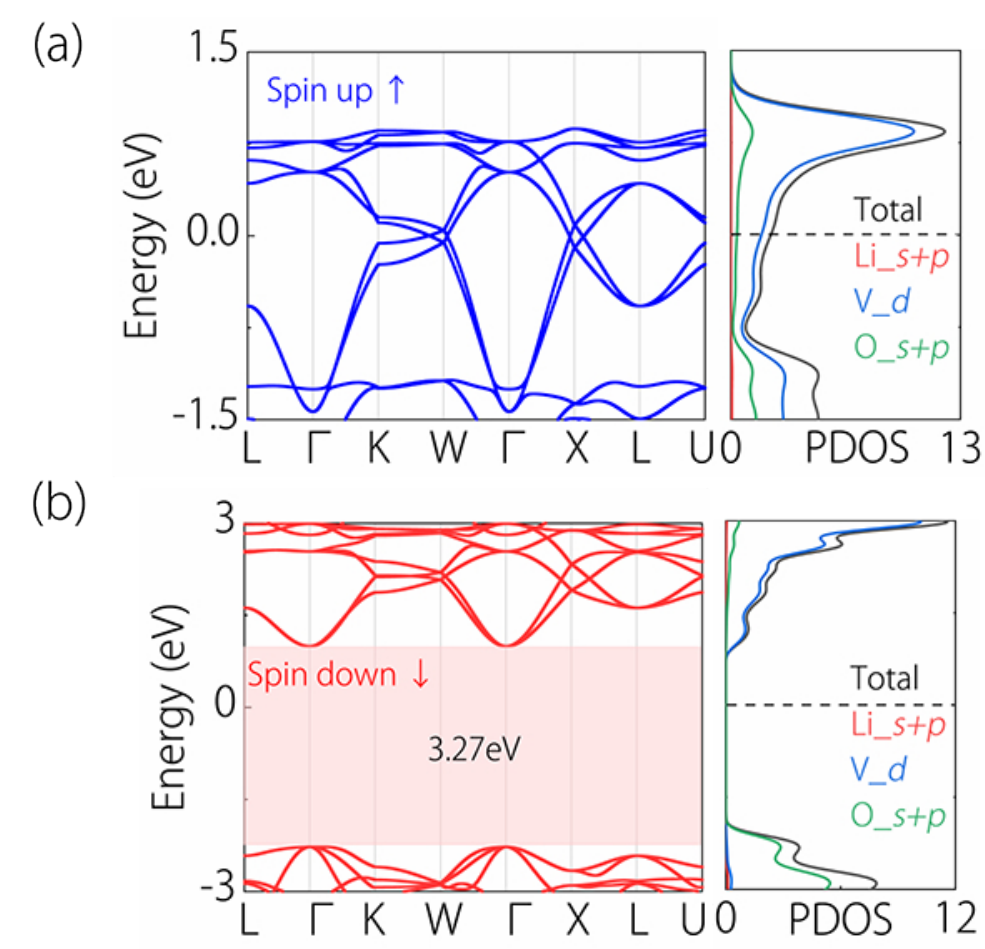}
\caption{The electronic band structures and projected density of states (PDOS) of LiV$_2$O$_4$ compound in the absence of SOC. (a) is for the spin-up ones, showing a metallic character with two bands crossing the Fermi level. (b) is for the spin-down channel, showing an insulating character with a big band gap of 3.27 eV.
\label{fig2}}
\end{figure}

Here we come to the band crossings in the spin-up channel. As displayed in Fig. 2(a), there are totally four band crossing points near the Fermi level, which happen in the K-W, W-$\Gamma$, $\Gamma$-X and X-L paths, respectively. After a careful scan of band structures, we find the crossings in the K-W, W-$\Gamma$ and $\Gamma$-X paths belong to a nodal loop in the $k_{z}$ = 0 plane. In Fig. 3(a), we show the enlarge view of orbital-component band structures in the W-$\Gamma$ and $\Gamma$-X paths. We can observe that, the bands with V-$d_{z^2}$ and V-$d_{x^2-y^2}$ orbital-component are inverted, which indicates the potential nontrivial band topology in LiV$_2$O$_4$ compound. These crossing points are not isolate but locate on a nodal loop in the $k_{z}$ = 0  plane. The profile of the nodal loop is shown in Fig. 3(b). This nodal loop is in fact under the protection of the glide mirror symmetry $G_{z}$: (x, y, z) $\to$ (x+1/4, y+3/4, -z+1/2). This requires the crossing bands possess opposite eigenvalues, which has been confirmed by our DFT calculations. Our calculations show the two crossing bands have the eigenvalue of +1 and -1, respectively. In the following, we denote this nodal loop as $NL_1$. Noticing the cubic symmetry of LiV$_2$O$_4$ compound, there also exist a nodal loop in both the $k_{x}$ = 0 and $k_{y}$ = 0 planes. These nodal loops cross with each other and form inner nodal chain structures, as shown in Fig. 3(e).

\begin{figure}
\includegraphics[width=8.8cm]{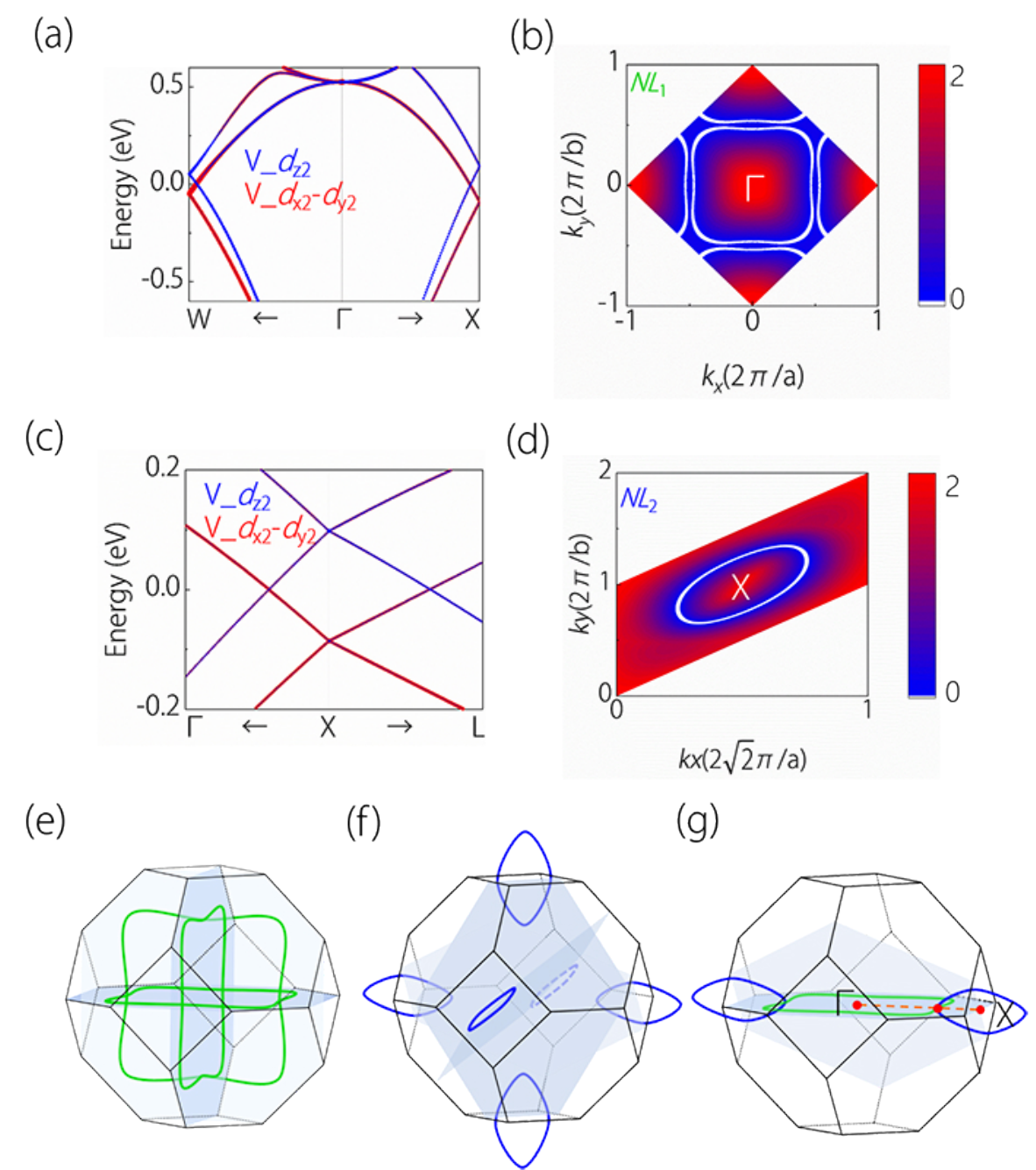}
\caption{(a) and (c) Orbital-projected band structure of LiV$_2$O$_4$, including V$-d_{z^2}$ orbital (blue), V-$d_{x^2-y^2}$ orbital (red). (b) and (d) showing the nodal loops in the $k_{z}$ = 0 plane and the $\Gamma$-X-L plane, which are labeled as $NL_1$ and $NL_2$, respectively. (e) and (f) are schematic illustration of the $NL_1$ and $NL_2$ in the Brillouin zone, respectively. (g) The schematic illustrations of the nodal chain in LiV$_2$O$_4$. The green and blue lines in (e)-(f) denote $NL_1$ and $NL_2$, respectively.
\label{fig3}}
\end{figure}

The band crossings in the $\Gamma$-X and X-L paths are also not isolate but belong to another nodal loop in the $\Gamma$-X-L plane. Figure 3(c) show the enlarged orbital-component band structures in the $\Gamma$-X and X-L paths. The profile of nodal loop in the $\Gamma$-X-L plane is shown in Fig. 3(d). Being similar with the nodal loop in the $k_{z}$ = 0 plane, this nodal loop is also protected by a glide mirror symmetry. The symmetry can be denoted as $G_{110}$: (x, y, z) $\to$ (-y+1/4, -x+3/4, z+1/2). We find the crossing bands have opposite mirror eigenvalues of $\pm$ 1. In the following, we denote this nodal loop as $NL_2$. According to the symmetry, there are totally three pairs of such nodal loop in the Brillouin zone, as shown in Fig. 3(f). Very interestingly, we find $NL_1$ and $NL_2$ are not isolate but share the same nodal point in the $\Gamma$-X path, thereby form the nodal chain structure. The profile of the nodal chain is shown in Fig. 3(g).

\begin{figure}
\includegraphics[width=8.8cm]{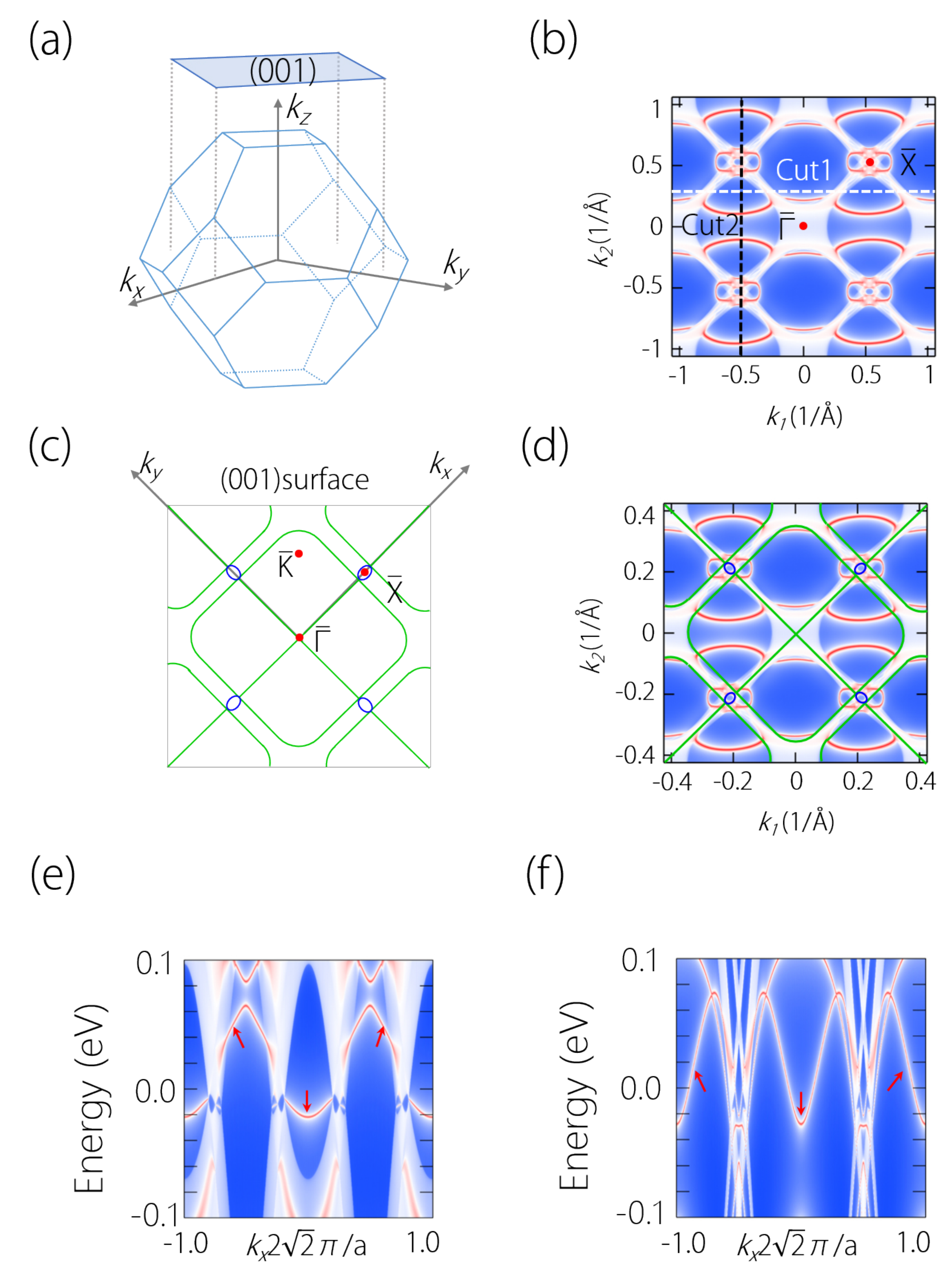}
\caption{(a) The bulk Brillouin zone and the projection onto the (001) surface. (b) Surface states on the (001) surface at the Fermi level. The sharp features are surface states, whereas the white region are projections of bulk bands. (c) Schematic illustration of the projection of nodal chain on the (001) surface. (d) (001) surface states with the profiles of projected nodal chain showing in the figure. (e)and (f) are surface band structures along the white and black cuts indicated in (b), where surface states connecting adjacent band crossings are observed.
\label{fig4}}
\end{figure}

Previously, nodal chain is mostly proposed in nonmagnetic materials including iridium tetrafluoride (IrF$_4$)~\cite{33}, WC-type HfC~\cite{39}, some hexagonal materials~\cite{41,73,74,75}, ternary Li$_2$XY (X = Ca, Ba; Y = Si, Ge) compounds~\cite{42}, metallic-mesh photonic crystal~\cite{43}, and carbon networks~\cite{40}. In these examples, the spin-polarization for the nodal-chain electrons is zero. Chang \emph{et al.} reported the first example of magnetic nodal chain in Heusler Co$_2$MnGa~\cite{34}. However, the electron states from both spin-up and spin-down channels involve together near the Fermi level, thus the conducting electrons near the nodal chain is only partially spin-polarized in Co$_2$MnGa. For this consideration, the nodal chain state in LiV$_2$O$_4$ compound is different with all the nodal chain materials proposed previously, because LiV$_2$O$_4$ is a half metal and the conducting electrons for the nodal chain is fully spin-polarized. Such nodal chain half metal is very potential in spintronics applications for high-speed information storage and processing.

Here we investigate the surface band structure of LiV$_2$O$_4$ compound. The equi-energy slice at the Fermi level in the (001) surface is shown in Fig. 4(b). We can observe several regions of drumhead surfaces states. To clarify the origin of these states, we show the profiles of nodal chain projected in the (001) surface. As shown in Fig. 4(c), loops $NL_1$ and $NL_2$ from the chain are shown in different colors. In Fig. 4(c), we map the profiles of projected nodal chain in the equi-energy slice. We find all the drumhead surfaces states originate from the nodal chain. Furthermore, we travel across the surface Brillouin zone along two typical paths [cut1 and cut2 in Fig. 4(b)], along which we expect to cross the nodal loops several times. Corresponding surface band structures are shown in Fig. 4(e) and (f). We indeed observe all surface states connect the adjacent band crossings, forming the $\omega$-shaped surface states. Such surface states are typical feature for nodal-chain fermion~\cite{34}.

\section{Effective model}

At the $\Gamma$ point, the symmetry is characterized by $C_{4v}$ point group. The generators can be chosen as \{$C_{4z}$, $M_{y}$, $C_{2z}$\}. One should note that in FM case, the spin-up and spin-down is decoupled in the absence of SOC. Each channel resembles an a spinless system. Therefore, the model has a time reversal symmetry for one channel with $\mathcal{T}$ = \textbf{\emph{K}}, \textbf{\emph{K}} is the complex conjugate. A minimal model for this two-band crossing around $\Gamma$ point can be generally written as
\begin{equation}
\mathcal{H}_\Gamma(\textbf{\emph{k}}) = \varepsilon_0(\textbf{\emph{k}})+\sum \limits_{i=x,y,z}d_{i}(\textbf{\emph{k}})\Sigma_{i},
\end{equation}
where $\varepsilon_0$(\textbf{\emph{k}}) is the overall energy shift, which can be neglected in our case. $\sigma$'s is the Pauli matrix, $d_i$(\textbf{\emph{k}}) is the functions of vector $\textbf{\emph{k}}$. The time reversal symmetry requires that
\begin{equation}
\mathcal{T}\mathcal{H}(\emph{\textbf{k}})\mathcal{T}^{-1} = \mathcal{H}(-\emph{\textbf{k}}),
\end{equation}
such that $d_y$ is an odd function of \textbf{\emph{k}} and $d_{x,z}$ is the even function of \textbf{\emph{k}}. Generally, one has d$_i$ in the form of
\begin{equation} \label{eqn2}
  \begin{split}
d_{x,z} = a_0 + a_1^{x,z}k_x^2 + a_2^{x,z}k_y^2 + a_3^{x,z}k_z^2 + \\
c_1^{x,z}k_xk_y + c_2^{x,z}k_yk_z + c_3^{x,z}k_xk_z,
  \end{split}
\end{equation}
\begin{equation}
d_{y} = b_1k_x +  b_2k_y + b_3k_z.
\end{equation}
For the mirror symmetry $M_y$ : ($x$,$y$,$z$) $\to$ ($x$,$-y$,$z$), one has
\begin{equation}
M_{y}\mathcal{H}(k_x,k_y,k_z)M_y^{-1} = \mathcal{H}(k_x,-k_y,k_z).
\end{equation}
Furthermore, from the DFT calculations, it is shown that the low energy states at the $\Gamma$ point belong to the following irreducible representations: \{$B_i$, $B_2$\}. The basis functions can be chosen as: \{$x^2$ - $y^2$, $xy$\}. Such that, $M_y$ = $\sigma_z$. This requires that $d_z$(\emph{\textbf{k}}) is an even function of $k_y$; terms linear to $k_yk_x$, $k_yk_z$ vanish. Then $d_z$(\emph{\textbf{k}}) = $a_0 + a_1^{z}k_x^2 + a_2^{z}k_y^2 + a_3^{z}k_z^2 + c_3^{z}k_xk_z$. And $d_{x,y}$(\emph{\textbf{k}}) are only the odd function of $k_y$. Therefore, $d_x$(\textbf{\emph{k}}) = $c_1^xk_xk_y$ + $c_2^xk_xk_y$  under the constrains from both $\mathcal{T}$ and $M_y$, and $d_y$ = $b_2k_y$.

In basis \{$B_i$, $B_2$\}, $C_{2z}$ : $(x,y,z)$ $\to$ $(-x, -y, z)$ is an identical matrix. It requires the Hamiltonian is an even function of $k_x$ and $k_y$, and Hamiltonian can be also proportional to $k_xk_y$. $d_y$ therefore vanishes in the presence of it. Additionally, term linear to $k_xk_z$ vanishes as well due to a negative sign arising from the $C_{2z}$ operation. Now, one has
\begin{equation}
d_{x} = c_1^xk_xk_y, d_{y} = 0, d_{z} = a_0+ a_1^zk_x^2 + a_2^zk_y^2 + a_3^zk_z^2.
\end{equation}

In this basis, $C_{4z}$ : $(x, y, z)$ $\to$ $(-y, x, z)$, it can be written as $C_{4z}$ = -$\sigma_0$. It requires coefficients of term which is odd function of $k_x$ and $k_y$ have opposite signs but the same absolute value. It also requires coefficients of term proportional to even order of $k_x$ and $k_y$ are same. Furthermore, the term $d_x$ = $c_1^xk_xk_y$ disappears due to a negative sign after operation. Therefore, the effective Hamiltonian at $\Gamma$ point can be written as
\begin{equation}
H_(\textbf{\emph{k}}) = [a_0 + a_1(k_x^2 + k_y^2) + a_3k_z^2]\sigma_z
\end{equation}
It shows a nodal loop on plane $k_z$ = 0.

Same discussion can be applied to the nodal loop on plane $k_y$ = 0 encircling the X point. The little group at the X point is $C_{2v}$. It has a two-fold rotation operation $C_{2z}$ and mirror operation $M_y$, and their combined operation $M_x$ : $(x, y, z)$ $\to$ $(-x, y, z)$. The general form of Hamiltonian can be written as
\begin{equation}
\mathcal{H}_M(\textbf{\emph{k}}) = \varepsilon_M(\textbf{\emph{k}})+\sum \limits_{i=x,y,z}f_{i}(\textbf{\emph{k}})\Sigma_{i}.
\end{equation}
Same argument is applied in the presence of $\mathcal{T}$ = \emph{\textbf{K}}. Hence, function $f_i$(\textbf{\emph{k}}) can be given by,
\begin{equation} \label{eqn2}
  \begin{split}
f_{x,z} = \alpha_0 + \alpha_1^{x,z}k_x^2 +  \alpha_2^{x,z}k_y^2 + \alpha_3^{x,z}k_z^2 + \\
 \gamma_1^{x,z}k_xk_y + \gamma_2^{x,z}k_yk_z+\gamma_3^{x,z}k_xk_z,
  \end{split}
\end{equation}
\begin{equation}
f_{y} = \beta_1k_x + \beta_2k_y + \beta_3k_z.
\end{equation}
The irreducible representations for this little group is $A_2$ and $B_2$, then the basis can be taken in the form of \{$xy$, $yz$\}.
Such that $C_{2z}$ =$\sigma_z$, which requires $f_3$(\textbf{\emph{k}}) is even function of $k_x$ and $k_y$, which can also be the function of $\gamma_1^zk_xk_y$. It asserts $f_{x,y}$ is odd function of $k_x$, $k_y$, such that
\begin{equation}
f_{x} = \gamma_1^xk_xk_z + \gamma_2^xk_yk_z,
\end{equation}
\begin{equation}
f_{y}= \beta_1k_x + \gamma_2k_y.
\end{equation}

In this basis, $M_y$ = - $\sigma_i$, it requires $f_i$(\textbf{\emph{k}}) is even function of $k_y$, such that considering all of generators.one has
\begin{equation}
f_{x}(\textbf{\textbf{k}}) = \gamma_1^xk_xk_z, f_{y}(\emph{\textbf{k}}) = \beta_1k_x,
\end{equation}
\begin{equation}
f_{z}(\emph{\textbf{k}}) = \alpha_0 + \alpha_1^zk_x^2 + \alpha_2^zk_y^2 + \alpha_3^zk_z^2.
\end{equation}
Finally, the Hamiltonian is
\begin{equation}
\mathcal{H} =(\alpha_0 + \alpha_1^zk_x^2 + \alpha_2^zk_y^2 + \alpha_3^zk_z^2) + \gamma k_xk_z\sigma_x + \beta k_x\sigma_y.
\end{equation}
It shows a nodal loop on plane $k_x$ = 0, the loop is described as
\begin{equation}
f_{z}(0,k_y,k_z) = \alpha_0 + \alpha_2k_y^2 + \alpha_3k_z^2.
\end{equation}
Till now, we have proved that there are two types of nodal loops in LiV$_2$O$_4$, which are shown in Fig. 3(e) and (f). The nodal loops touch with each other at one point along the $\Gamma$-M direction and form the nodal chain structure in the Brillouin zone, as shown in Fig. 3(g).

\section{Robustness of nodal chain and SOC effect}

To ensure the fully spin-polarized nodal chain, two conditions need to be considered: first, it requires a band gap in the spin-up channel, which enables the half metallic character; second, the band crossings in the spin-down channel are necessary, which ensure the presence of nodal chain.

\begin{figure}
\includegraphics[width=8.8cm]{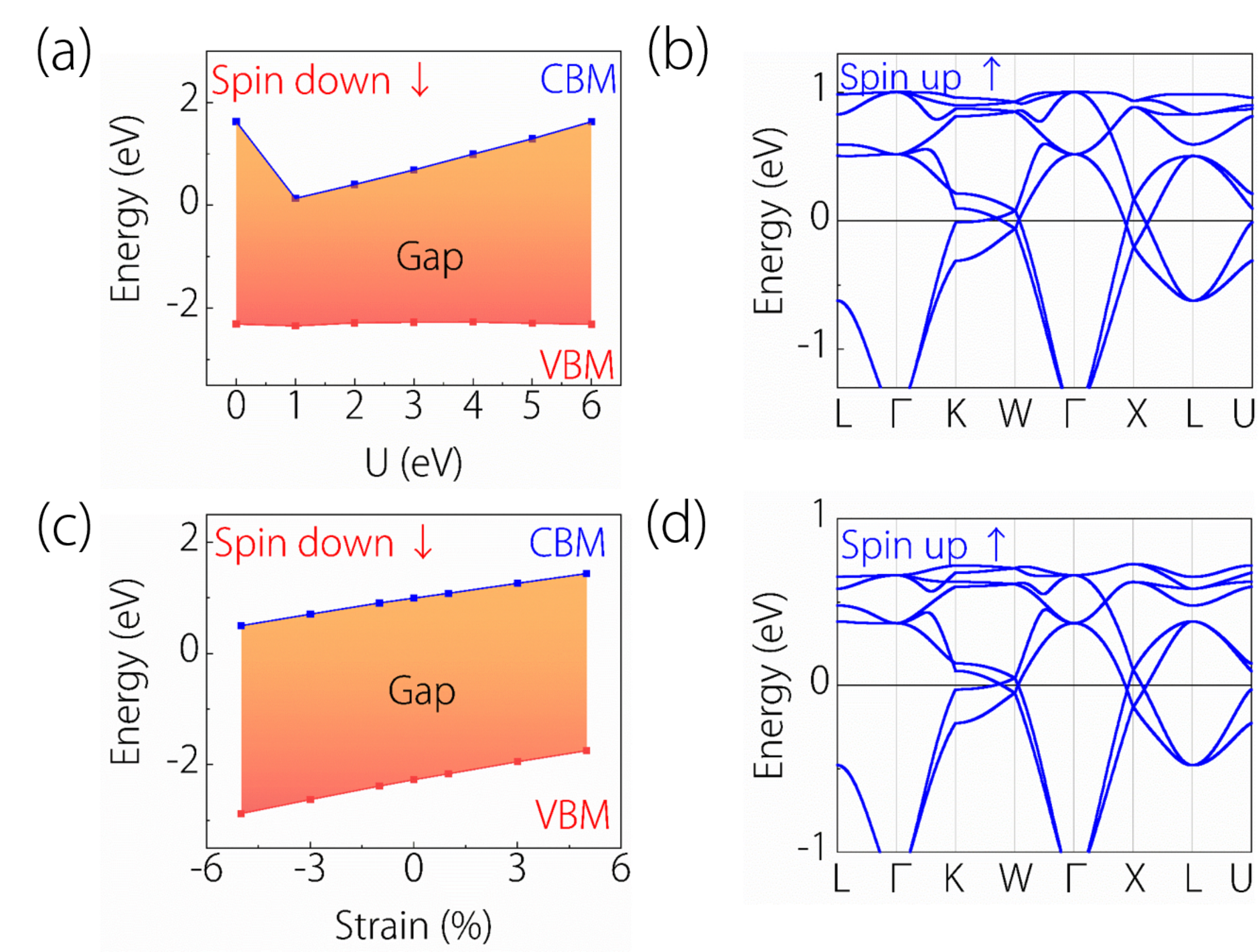}
\caption{(a) In the spin-down channel, the curves of the valence band maximum (VBM) and the conduction band minimum (CBM) under different \emph{U} values. The area formed by the curve showing the band gap in the spin-down channel. (b) The band structure of LiV$_2$O$_4$ compound in the spin-up channel with \emph{U} at 6 eV. (c) The curves of VBM and CBM under different strains in the spin-down channel. (d) The spin-up band structure under a 5$\%$ tensile strain.
\label{fig5}}
\end{figure}

To show the robustness of the spin-polarized nodal chain in LiV$_2$O$_4$ compound, we have examined the electronic band structure against the electron correlation effects and lattice strains. In Fig. 5(a), we show the positions at the bottom of conduction band and at the top of valence band versus the \emph{U} values of V element. We can find that, the band gap between the conduction band and valence band in the spin-down channel always exists with the \emph{U} values shifting from 0 to 6 eV. For the spin-up channel, the conduction and valence bands cross with each other. We find such band crossings always retain during the shift of \emph{U} values. In Fig. 5(b), we show the band structure of LiV$_2$O$_4$ compound in the spin-up channel with \emph{U} at 6 eV. The band crossings produce the fully spin-polarized nodal chain. Moreover, we have also investigated the electronic band structure under hydrostatic strains. As shown in Fig. 5(c), we find the band gap in the spin-down channel can retain under $\pm$ 5$\%$ hydrostatic strains (where ``+" represents tensile strain and ``-" represents compressive strain). Meanwhile, the band crossings in the spin-up channel can also retain in the period. Figure 5(d) shows the band structure in the spin-up channel under a 5$\%$ tensile strain, where band crossings for nodal chain are observed. These results suggest the fully spin-polarized nodal chain in LiV$_2$O$_4$ compound is very robust, which can be meaningful for its future detections in experiments.

Finally, we discuss the SOC effect on the electronic band structure. The resulting band structure under SOC is shown in Fig. 6(a). We can find that, the bands from both spin channels conjunct together under SOC, but the band details do not change much near the Fermi level. As has been discussed above, $NL_1$ and $NL_2$ of the nodal chain are protected by specific glide mirror symmetries. In LiV$_2$O$_4$ compound, the ground magnetic moment ordering is along in the [001] direction. All the glide mirror symmetries except $G_z$ will be broken under such a magnetization direction. As the results, most of the loops for nodal chain will be gapped under SOC. However, one of $NL_1$ in the $k_z$ = 0 plane would retain because the crossing bands still have opposite $G_z$ eigenvalues ($\pm$ i), as protected by the glide mirror symmetry $G_z$. These arguments have been verified by our DFT calculations. As shown by the enlarged band structures in Fig. 6(b), we can find that the band crossings in the K-W, W-$\Gamma$ and $\Gamma$-X paths are not gapped under SOC but that in the X-L path is gapped with gap size of $\sim$ 14 meV. These results have shown that, LiV$_2$O$_4$ compound shows a single nodal loop under SOC, as displayed in Fig. 6(c).

\begin{figure}
\includegraphics[width=8.8cm]{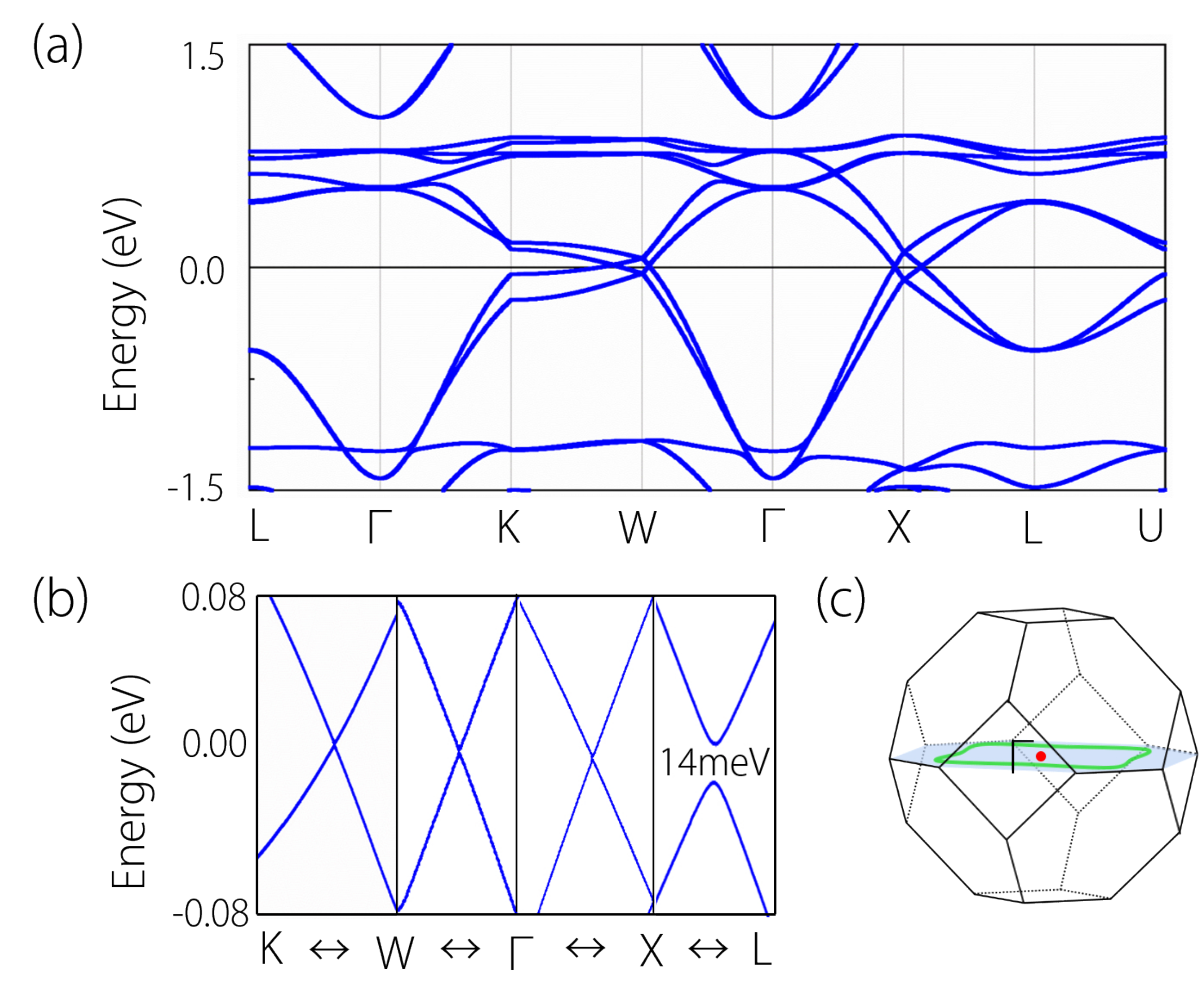}
\caption{(a) Electronic band structure of LiV$_2$O$_4$ under SOC with magnetization along the [001] direction. (b) The enlarged band structure in the K-W, W-$\Gamma$ and $\Gamma$-X paths are not gapped under SOC, whereas that in the the X-L path is gapped with the gap size of 14 meV. (c) Schematic illustration of a single nodal loop under SOC in the $k_z$ = 0 plane under the [001] magnetization.
\label{fig6}}
\end{figure}

\section{CONCLUSION}

In conclusion, we have demonstrated the presence of fully spin-polarized nodal chain in an existing material LiV$_2$O$_4$. The material shows a ferromagnetic ground state with the spin ordering in the [001] direction. It manifests a half metal band structure with a metallic character in the spin-up channel but an insulating one in the spin-down channel. The band crossings near the Fermi level form two types of Weyl loops, which conjunct with each other at a specific point. This gives rise to the formation of nodal chain. The nodal chain exists only in the spin-up channel; hence, it is fully spin-polarized. We find the nodal chain show $\omega$-shaped surface states, which are also fully spin-polarized. We further find the nodal chain is very robust against the electron correlation effects and the lattice strain. This work provides an excellent platform to investigate fully spin-polarized nodal-chain fermions in realistic materials, as well as bring promising applications in spintronics.

\begin{acknowledgments}
This work is supported by National Natural Science Foundation of China (Grants No. 11904074), Nature Science Foundation of Hebei Province (No. E2019202222 and E2019202107). One of the authors (X.M. Zhang) acknowledges the financial support from Young Elite Scientists Sponsorship Program by Tianjin.
\end{acknowledgments}

\end{document}